\documentclass[%
 reprint,
 amsmath,amssymb,
 aps,
]{revtex4-2}

\pdfoutput=1

\usepackage{graphicx}
\usepackage{dcolumn}
\usepackage{bm}
\usepackage{amssymb}
\usepackage{amsmath}
\usepackage{commath}
\usepackage{graphicx,bm}
\usepackage{verbatim}
\usepackage[caption=false]{subfig}

\usepackage{color}
    \definecolor{darkgreen}{rgb}{0,0.5,0}
    \definecolor{darkred}{rgb}{0.5,0,0}
    \definecolor{darkblue}{rgb}{0,0,0.6}
    \definecolor{purple}{rgb}{0.4,.2,0.7}

\usepackage[%
  breaklinks=true,
  colorlinks=true,
  linkcolor=blue,
  urlcolor=blue,
  citecolor=blue
]{hyperref}

\usepackage{xr}

\newcommand{\bigO}[1]{\ensuremath{\mathcal{O}(#1)}}
\newcommand{\inv}[1]{\frac{1}{#1}}

\newcommand*\diff{\mathop{}\!\mathrm{d}}

\newcommand\pder[2][]{\ensuremath{\frac{\partial#1}{\partial#2}}}
\newcommand*\LR{{}^L_R}
\newcommand*\HplusLR{\mathcal{H}^+_{\LR}}

\begin{document}

\preprint{APS/123-QED}

\title{Charged Static AdS Black Hole Binaries}

\author{William~D.~Biggs}
\author{Jorge~E.~Santos}%
\affiliation{%
 Department of Applied Mathematics and Theoretical Physics, University of Cambridge, Cambridge, CB3 0WA, United Kingdom
}%

\date{\today}

\begin{abstract}
We construct the first binary black hole solutions of Einstein-Maxwell theory in asymptotically anti-de Sitter space. The attractive force between the two black holes is balanced by the addition of a background electric field, sourced at the conformal boundary. There is a continuous family of bulk solutions for a given boundary profile and temperature, suggesting there is continuous non-uniqueness. We investigate the charges of the solutions and verify numerically that they satisfy a first law of black hole mechanics relation.
\end{abstract}

\maketitle

\paragraph*{Introduction.}
Very few stationary multi-black hole solutions are known to exist. Indeed, stationary black holes are often hypothesised to be uniquely defined in terms of a small number of asymptotic properties, such as their energy, charge and angular momentum, which would forbid the existence of multi-horizon solutions. This hypothesis, generally referred to as the ``no-hair theorem'', has been proven to hold under certain assumptions \cite{Israel:1967wq, Israel:1967za, Penney:1968.174.1578, Chase:1970omy,Bekenstein:1972.28.452, Bekenstein:1972.5.1239, Bekenstein:1972.5.2403, Teitelboim:1972.5.2941, Robinson:1977aa, Bunting:1987bb,Heusler:1996ft, Bekenstein:1995un, Bekenstein:1996pn, Sudarsky:1995zg}.

However, the no-hair theorem is known to be violated in various other circumstances. For example, there exist black hole solutions with matter hair \cite{Gubser:2008px, Hartnoll:2008kx, Dias:2011at, Volkov:1998cc, Herdeiro:2014goa, Herdeiro:2016tmi}, and in higher dimensions, black hole horizons can have non-spherical topology \cite{Emparan:2001wn, Figueras:2014dta}.

Solutions containing multiple black holes also exist, in each of which the gravitational attraction between the black holes is balanced by some other force. In Einstein-Maxwell theory, the famous Majumdar-Papapetrou solution \cite{Majumdar:1947eu, Papapetrou:1947b} is a configuration of four-dimensional black holes with extremal charge in which their electric repulsion balances
their gravitational attraction. Likewise, in Einstein-Maxwell theory or Einstein-Maxwell-Dilaton theory, an external magnetic field can be used to balance oppositely charged extremal black holes \cite{Emparan:1999au}. Similar exact systems of extremal black holes are also known in higher dimensions \cite{Myers:1986rx, Ishihara:2006iv}. Indeed, there are many more multi-black hole solutions known to exist in higher dimensions, some supported by the centrifugal force arising from rotation \cite{Elvang:2007rd, Iguchi:2007is, Izumi:2007qx, Elvang:2007hs}, and others \cite{Elvang:2002br, Tomizawa:2007mz, Iguchi:2007xs, Astorino:2022fge, Tomizawa:2024tkh} supported by gravitational solitons known as ``bubbles of nothing" \cite{Witten:1981gj}. Finally, stationary black binaries with de Sitter asymptotics have also been obtained in pure gravity \cite{Dias:2023rde, Dias:2024dxg}, this time being held apart by the expansion due to a positive cosmological constant.

In this letter we turn our attention to the question of whether binary solutions can exist in anti-de Sitter (AdS) space. We wish these solutions to be asymptotically globally AdS, tending towards the geometry of global AdS space, which is described by the metric:
\begin{subequations}
\begin{equation}
    \label{eqn:AdSf}
    \diff s^2_{\rm AdS} = - f(R) \diff t^2+f(R)^{-1} \diff R^2+R^2 \diff \Omega^2_{d-2},
\end{equation}
with \vspace{-15pt}
\begin{equation}
    f(R) = 1+\frac{R^2}{\ell_d^2}.
\end{equation}
\end{subequations}
In these coordinates, the conformal boundary is situated at $R \to \infty$, and $\ell_d$ denotes the AdS length scale in $d$ dimensions (here, and throughout, $d$ is reserved to denote the number of bulk spacetime dimensions). 

\begin{figure}[tb]
    \centering
    \includegraphics[width=0.3\textwidth]{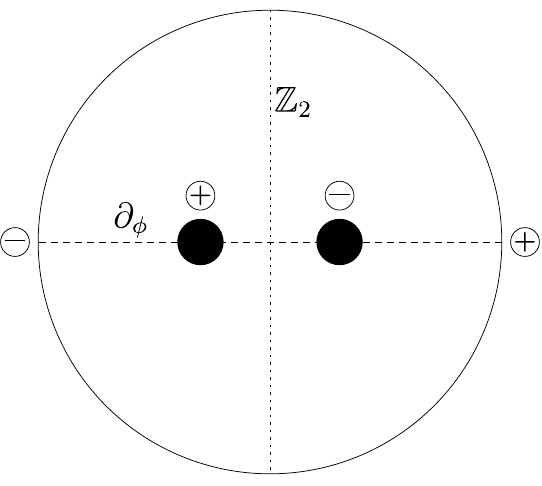}
    \caption{A schematic drawing of a symmetric binary solution in AdS with an $\ell = 1$ boundary profile. The vertical, dotted line is a plane of reflective symmetry, whilst the horizontal, dashed line is an axis of rotation. The plus and minus symbols represent the charges of the horizons and the poles of the conformal boundary. The system is in static equilibrium since the two black holes are attracted to opposite sides of the boundary, balancing their mutual attraction.}
    \label{fig:sketch}
\end{figure}

The presence of a negative cosmological constant contributes an additional gravitational potential well (compared with asymptotically flat space), and so one may expect that binaries cannot exist in Einstein-Maxwell theory, since it seems even electrically extremal black holes wouldn't have sufficient electric repulsion to balance the gravitational attraction. However, in AdS there is also the possibility to add a background electric field which is non-zero at the conformal boundary. We will show that the addition of such a electric potential allows for the existence of static, charged, binary black hole solutions with AdS asymptotics in both four and five dimensions. Fig.\ref{fig:sketch} gives the sketch of an indicative example of such a system, in which the boundary electric field is positive on one pole of the boundary sphere, and negative on the other. Then two oppositely charged black holes are attracted to opposite sides of the conformal boundary, leading to the possibility of a binary system.

Such solutions have an additional source of interest due to the AdS/CFT correspondence \cite{Maldacena:1997re, Witten:1998qj, Aharony:1999ti}, which is a duality between a theory containing a strongly coupled conformal field theory (CFT) and a gravitational theory in AdS space in one higher dimension. 

On the field theory side, adding a background electric field in the bulk corresponds to deforming the CFT theory by the addition of a chemical potential, which acts as an electric source for the theory, given by
\begin{equation}\label{eqn:source}
    A^{(0)} = \mu(\theta)\diff t,
\end{equation}
where $\theta$ is the polar angle of the sphere, and $\mu(\theta)$ is a profile which we are free to choose. This chemical potential excites a Maxwell field in the bulk theory, with its leading order behaviour at infinity dictated by the profile, $\mu(\theta)$.
As we will discuss, this allows for the existence of binary solutions, which correspond holographically to states of the CFT  under the influence of this electric source residing on a $(d-1)$-dimensional Einstein static Universe (ESU) of radius $\ell_d$.
In Ref.~\cite{Costa:2015gol}, other bulk solutions with the same boundary conditions were investigated, one, a soliton with no horizon, and another, a single black hole whose horizon is polarized due to the background electric field. Similar solutions have also been studied in the context of more complex boundary electric fields in \cite{Herdeiro:2015vaa,Herdeiro:2016plq}.

Interestingly, we find that for fixed choices of the temperature and the amplitude, $\mu_1$, of the electric field, there is a \emph{two-dimensional} solution space of binaries, with the two continuous parameters corresponding to a local chemical potential of each of the black hole horizon. Only a one-dimensional sub-family of this space preserves the $\mathbb{Z}_2$ symmetry of the boundary data.
\paragraph*{Numerical Construction.}
The black hole binaries are solutions to the Einstein-Maxwell equations possessing a Killing vector field, $\partial_t$. Moreover, let us assume axisymmetry in four dimensions and $SO(3)$ symmetry in five dimensions, so  that the solutions are cohomogeneity-two. For the asymptotic behaviour, we require that the solutions are asymptotic to global AdS, \eqref{eqn:AdSf}, and we will impose a non-zero profile for the gauge field at infinity, given by Eq.~\eqref{eqn:source}. This latter condition will excite a Maxwell field in the bulk. We enforce the black holes are electrically charged, by allowing only the $t$-component of the gauge potential to be non-zero.

In the Supplementary Material we fully explain the numerical method (based on techniques more generally explained in Ref. \cite{Dias:2015nua}) used to attain stationary solutions to the Einstein-Maxwell equations with such boundary conditions, but let us here briefly summarise the key challenges and how they are overcome. 

Firstly, the Einstein equation yields partial differential equations (PDEs) of mixed hyperbolic-elliptic character unless an appropriate gauge is chosen. To overcome this issue, we use the DeTurck trick (introduced in Ref. \cite{Headrick:2009pv} and reviewed in Refs. \cite{Dias:2015nua,Wiseman:2011by}), which adds a term to the Einstein equation to make it explicitly elliptic for the symmetry class of interest. This yields the Einstein-DeTurck equation
\begin{subequations}
\begin{equation}\label{eqn:EDT}
    0 = R_{ab} + \frac{d-1}{\ell_d^2}g_{ab} - 2 \Tilde{T}_{ab} - \nabla_{(a}\xi_{b)}
\end{equation}
where $\Tilde{T}_{ab}$ is the trace-reversed stress tensor given by
\begin{equation}
    \Tilde{T}_{ab} = F_a{}^cF_{bc}-\inv{2(d-2)}g_{ab}F_{cd}F^{cd}
\end{equation}
and the AdS length scale, $\ell_d$, is related to the cosmological constant, $\Lambda$, by $\Lambda = -(d-1)(d-2)/(2\ell_d^2)$. The vector, $\xi^a$, is called the \textit{DeTurck vector}, and is defined by
\begin{equation}
\xi^a = g^{cd}\left[\Gamma^a_{cd}(g)-\Gamma^a_{cd}(\Bar{g})\right]
\end{equation}
where $\Gamma^a_{cd}(g)$ is the Christoffel connection associated to a metric, and $\Bar{g}_{ab}$ is a reference metric which we are free to choose. Eq.~\eqref{eqn:EDT} must be solved alongside the Maxwell equation,
\begin{equation}
    0 = \nabla^a F_{ab},
\end{equation}
\end{subequations}
which is already elliptic since the gauge vector is assumed to be proportional to the static Killing vector field. 

Only solutions to \eqref{eqn:EDT} with $\xi = 0$ are solutions to the Einstein equation; otherwise, they are called \textit{Ricci solitons}. Due to the presence of the Maxwell field, our current scenario does not satisfy the assumptions of any proof forbidding the existence of Ricci solitons \cite{Figueras:2011va, Figueras:2016nmo}. 
However, the fact that the equations are elliptic implies that no solution with non-zero $\xi^a$ can be arbitrarily close to a solution to Einstein's equation, meaning that tracking the value of the norm, $\xi^a \xi_a$, provides a useful convergence test, which is presented in the Supplementary Material.

Another difficulty is that the solutions do not naturally live on a rectangular integration domain, and so patching techniques are necessary. The use of patching also allows for the use of two different sets of coordinates in different regions of the domain. Near the black holes, we use ring-like coordinates, well-adapted for the presence of a binary system and based on the Israel-Khan solution \cite{Israel:1964a}, a binary solution in asymptotically flat space in which the black holes are held apart by a conical strut.
On the other hand coordinates similar to those of empty, global AdS are used in the asymptotic region. Though the Israel-Khan metric contains this conical singularity, as explained in the Supplementary Material, the AdS binary solutions obtained do not, due to a warping factor being added to the Israel-Khan metric when building a reference metric.
\paragraph*{Results.}
We were able to obtain a very large family of binary solutions. Let us focus \footnote{We were also able to obtain solutions with a more general boundary profile, $\mu(\theta) = \mu_1 \cos \theta + \mu_2 \cos 2\theta$, which possesses no symmetry about $\theta = \pi/2$.}
on the case in which the boundary profile for the chemical potential is given by an $\ell = 1$ mode so that we have
\begin{equation}\label{eqn:l=1}
\mu(\theta) = \mu_1 \cos\theta,
\end{equation}
and, for the remainder of the letter, let us set $\ell_d = 1$. We will use $L$ and $R$ subscripts to differentiate between quantities associated to the left (on the $\theta = \pi$ side) and right (on the $\theta = 0$ side) horizons, respectively. The horizons have temperatures, $T_L$ and $T_R$. We assumed that the two temperatures were equal, with $T_L = T_R =: T$, though it seems likely that they could take different values. 

From the bulk solution, one can extract the dual holographic stress tensor, $\langle T_{\mu\nu} \rangle$, and conserved current, $\langle J^\mu \rangle$ of the boundary field theory via the standard process of holographic renormalization \cite{deHaro:2000vlm, Dias:2022eyq} using Fefferman-Graham gauge \cite{Fefferman:1985}. We explicitly carry out this procedure in the Supplementary Material. Of particular interest will be the charge density, $\rho \equiv \langle J^t \rangle$, and the total energy,~$E$, which can be defined as an integral of $\langle T^t{}_t \rangle$ over a Cauchy slice, $\Sigma$, of the boundary.

The total charge is given by
\begin{equation}
Q \equiv \int_{\Sigma} \sqrt{h} \rho(\theta) = \inv{4\pi}\int_{\Sigma} \star F,
\end{equation}
where $h$ is the determinant of the induced metric of $\Sigma$ and the second equality follows from the definition of the charge density, $\rho$. We shall also consider the integral of the charge density over the left and right hemispheres, respectively denoted by $Q_L^{(inf)}$ and $Q_R^{(inf)}$.
 
The charge of each horizon can be defined by a surface integral over a spatial cross-section of the horizon:
\begin{equation}
    Q^{({\rm hor})}_{\LR} = \inv{4\pi} \int_{\HplusLR} \star F.
\end{equation}
Gauss' law implies that $Q^{({\rm hor})}_L + Q^{({\rm hor})}_R = Q$. 
As usual, the entropy, $S_{\LR}$, of each horizon is proportional to its area.

Interestingly, even for a given choice of $\mu_1$ and $T$, there are still further parameters which must be fixed to specify a particular binary solution. Due to the fact each horizon is a Killing horizon with Killing vector field, $\partial_t$, the vector potential is constant on each each horizon:
\begin{equation}
    A\big{|}_{\mathcal{H}^+_{\LR}} = \nu_{\LR} \diff t,
\end{equation}
where $\nu_L$ and $\nu_R$ are two constants, which we dub the \textit{local chemical potential} at the respective horizons. It turns out $\nu_L$ and $\nu_R$ need not be zero. Indeed, for fixed values of $\mu_1$ and $T$ there is a continuous range of values of $\nu_L$ and $\nu_R$ for which we were able to find binary solutions.

Note that there is no gauge transformation which can alter $\nu_L$ and $\nu_R$ whilst remaining in the static gauge (with $A \propto \diff t$ and $A_t$ time-independent) except shifting the vector potential by a constant everywhere. Hence fixing that the $\ell = 0$ mode of the boundary profile is zero fixes the gauge of the vector potential completely.

From the perspective of AdS/CFT, the presence of the $\nu_L$ and $\nu_R$ parameters is extremely surprising. 
Each binary solution corresponds to a state of the CFT under the influence of the chemical potential, and hence, there is a whole family of stationary CFT states parameterised by these continuous parameters, even for a fixed value of the temperature. Since the parameters $\nu_L$ and $\nu_R$ are defined by the behaviour of the solution on the horizons, which lie very deep within the bulk, it is not at all clear what these parameters correspond to in the field theory.

\begin{figure}[t!]
    \includegraphics[width=\linewidth]{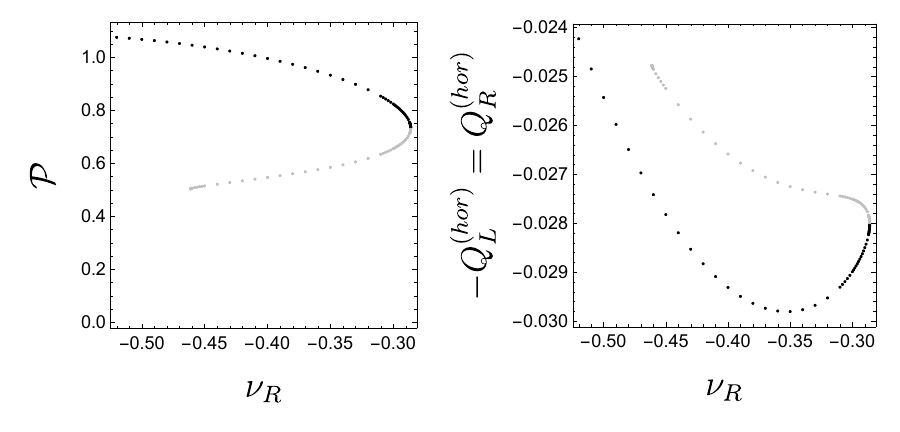}
    \caption{The proper distance (left) between and the electric charges (right) of the horizons for each of the $\mathbb{Z}_2$-symmetric solutions with $T=1$ and $\mu_1 = 1.5$ in four dimensions. There is a turning point at $\nu_R \simeq -0.286$, either side of which we use different shades to indicate the two branches of solutions.}
    \label{fig:turning}
\end{figure}

In general, the  $\mathbb{Z}_2$ symmetry, about $\theta = \pi/2$, of the boundary data is broken in the bulk. However, one can force the bulk solution to inherit this symmetry by setting $\nu_L = -\nu_R$. In Fig.~\ref{fig:turning}, we plot the value of the proper distance between the horizons on the left and the value of the charges on the right for each of the $\mathbb{Z}_2$-symmetric solutions with $\mu=1.5$ and $T = 1$.
These necessarily satisfy $S_L = S_R$ and $Q^{({\rm hor})}_L = -Q^{({\rm hor})}_R$, thus, though the charge of each horizon depends on $\nu_R$, the \textit{net charge} is always zero. Therefore, the existence of this one-dimensional family of $\mathbb{Z}_2$-symmetric solutions represent continuous non-uniqueness of the bulk solution for given boundary data. 

There is a turning point for $\nu_R$, at $\nu_R \simeq -0.286$. Along either branch, which are shown in two different shades, solutions cease to exist as you increase the parameter beyond this value of $\nu_R$, however, since no singularities arise and some physical quantities begin to have large derivatives with respect to $\nu_R$, it is highly suggestive that new solutions (the other branch) will be able to be found by instead decreasing $\nu_R$ once again.
We used the method described in Sec.~VII-B of Ref.~\cite{Dias:2015nua} to perturb from a solution on one branch to a solution on the other.

\begin{figure}[t]
    \centering
    \includegraphics[width=0.48\textwidth]{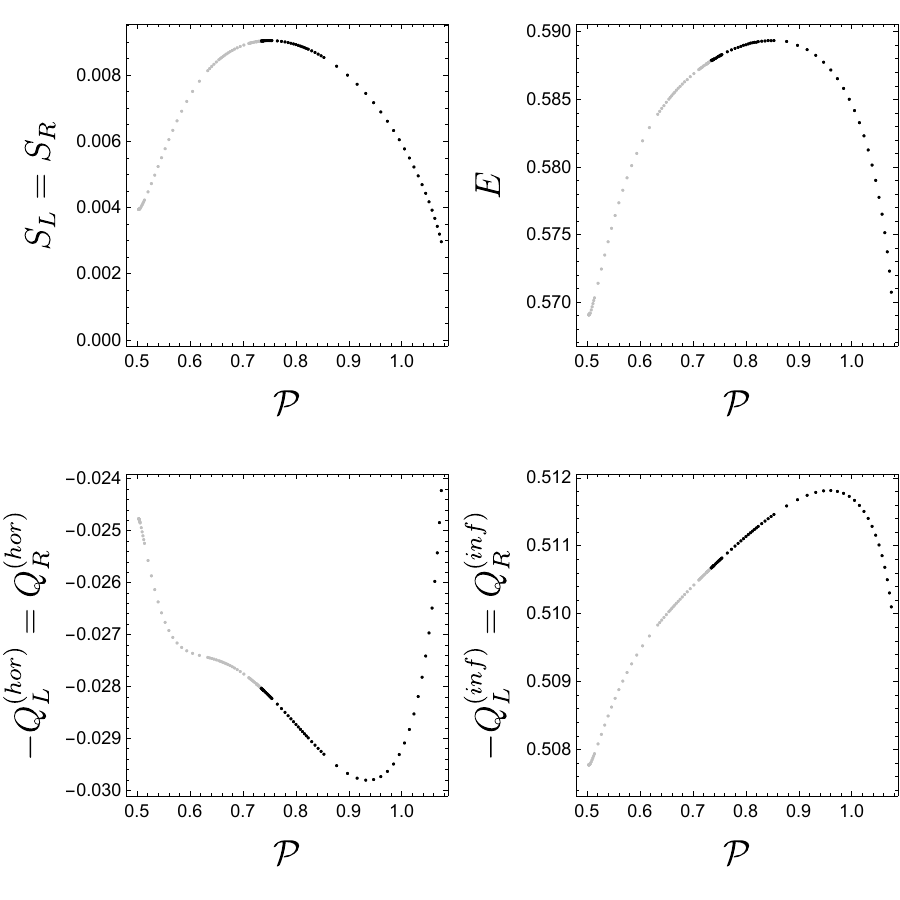}
    \caption{A plot of various thermodynamic quantities (clockwise from bottom-left: charges of the horizons, entropy, energy, charges of the asymptotic hemispheres) against the proper distance between the horizons for four-dimensional $\mathbb{Z}_2$-symmetric solutions with the same temperature and boundary potential, $T=1$ and $\mu=1.5$. The differently shaded points correspond to different branches, either side of the turning point. 
    }
    \vspace{-5pt}
    \label{fig:SEQQ}
\end{figure}

In Fig.~\ref{fig:SEQQ}, we plot how the entropy, energy, charge of the horizon and charge of the boundary hemispheres of the solutions (again with $T=1$ and $\mu_1 = 1.5$) depend on the proper distance between the horizons.
This plot helps to elucidate the necessity of the extra continuous parameter, $\nu_R$. One can alter the charges of the black holes, whilst in a counteracting fashion changing the distance between them, thereby maintaining a static configuration. From the bottom-left panel, we see that the balancing act between charge and proper distance does not occur in a simple monotonic fashion, due to the complicated balancing of multiple forces.

We see in the top-right panel of Fig.~\ref{fig:SEQQ} that the energy takes a maximum value, at $\mathcal{P} \simeq 0.85$. At this point, to leading order, the distance between the horizons is changing, but the energy is unaltered. This suggests there is a zero-mode, and thus a change of stability. Since the entropy decreases at this value of $\mathcal{P}$, it suggests that the solutions with $\mathcal{P}<0.85$ are more stable than those with $\mathcal{P}>0.85$ within this subfamily of solutions. We similarly find evidence of a turning point for the energy among subfamilies of the five-dimensional solutions with fixed temperature and boundary chemical potential.

As seen in Fig.~\ref{fig:turning}, $\nu_R$ can only be decreased to certain values on either branch before solutions cease to exist. The Kretschmann scalar at the horizons increases as $\nu_R$ decreases towards these values, hence it is possible that singularities arise at these points in the parameter space. However, the evidence for this is inconclusive and solutions could cease to exist without the presence of singularities (perhaps in a similar manner to Ref.~\cite{Dias:2021vve}) or there could be further turning points.

The binary solutions satisfy a first law of black hole mechanics, or thermodynamics, relationship, given by
\begin{equation}\label{eqn:firstE}
    \delta E = \hspace{-7pt}\sum_{i \in \{L,R\}} \hspace{-7pt}\left(T_i \delta S_i + \nu_i \delta Q^{({\rm hor})}_i \right) + \int_\Sigma \diff \bm{x} \sqrt{h} \mu(\theta)  \delta \rho(\theta).
\end{equation}
The final term must be calculated as an integral due to the fact that the boundary chemical potential, $\mu(\theta)$, is not constant. However, if one is only interested in variations which keep $\mu(\theta)$ fixed, one can instead consider
\begin{equation}
    G = E - \hspace{-7pt}\sum_{i \in \{L,R\}} \hspace{-7pt} \left(T_i  S_i + \nu_i Q^{({\rm hor})}_i \right) -\int_\Sigma \diff \bm{x} \sqrt{h} \mu(\theta)  \rho(\theta),
\end{equation}
which will consequently satisfy a first law relationship,
\begin{equation}\label{eqn:firstG}
    \delta G = - \sum_{i \in \{L,R\}} \left(S_i \delta T_i + Q^{({\rm hor})}_i \delta \nu_i \right),
\end{equation}
under such variations. The fact that only variations of the charges, rather than any quantities within an integral, are involved in \eqref{eqn:firstG} makes it easier to verify on the numerical solutions, and we were able to do so for the binary solutions in both four and five dimensions.

Under the AdS/CFT duality, the first law described in Eq.~\eqref{eqn:firstE} should also relate variations of charges of the field theory which should be possible to be defined without reference to a gravitational bulk. However, whilst $\nu_{i}$ and $Q^{({\rm hor})}_i$ have clear physical meanings in the bulk as quantities defined on the horizon, their meaning in terms of the boundary theory is not at all clear. Therefore the presence of the terms in the first law concerning these quantities is perplexing from the perspective of the field theory. The situation is somewhat analogous to the thermodynamics of the asymptotically flat five-dimensional black rings discovered in \cite{Emparan:2004wy}, where dipole charges do appear in the first law of black hole mechanics \cite{Copsey:2005se}, and yet cannot be measured at spatial infinity.

Furthermore, it is not clear that $G$ is the value one would derive as the free energy by computing directly from the action, and indeed it seems unlikely that there even is a meaningful definition of the free energy at all. This is because Wick rotating the bulk solutions yields a Euclidean solution that is singular at the horizons unless the vector potential vanishes there, \textit{i.e.} $\nu_R = \nu_L = 0 $. No gauge transformation can be taken to ensure these conditions unless $\nu_R = \nu_L$, which is generically not the case. The binary solutions are static, but intrinsically Lorentzian in nature.
\paragraph*{Discussion.}
In this letter we have presented the first examples of binary black hole solutions in Einstein-Maxwell theory with a negative cosmological constant in both four and five dimensions.\footnote{We note though that multi-black hole solutions with anti-de Sitter boundary conditions have been \emph{conjectured} to exist in Ref.~\cite{Anninos:2013mfa} within the context of the bosonic sector of Fayet-Iliopoulos $\mathcal{N}=2$ gauged supergravity with a cubic prepotential. Indeed, the same reference discusses some of the properties of these intriguing multi-black hole solutions using the probe approximation.} We have obtained a large family of such solutions, and in particular we showed that there are unexpected, extra parameters which are given by the value of the gauge potential at each of the horizons but do not have clear meanings from the boundary.

One source for future research would be to extend the solution space. It would be particularly interesting to ascertain whether the magnitude of the chemical potential can be taken to be arbitrarily small whilst still allowing for the existence of binary solutions. Moreover, if multi-black hole solutions exist near extremality in $d\geq5$, they could provide a metastable phase in the RG flow, as conjectured in Ref.~\cite{Horowitz:2022leb}, possibly establishing a connection with the fragmentation scenario described in Ref.~\cite{Maldacena:1998uz}.


One could also allow for rotation of the binary black holes. This could be done in a way still respecting axisymmetry, by having the black holes both rotate on their shared axis, à la Ref.~\cite{Dias:2024dxg}. This would introduce another degree of non-uniqueness, since the net angular momentum would still be zero if the black holes were spinning in opposite directions at the same speed. Moreover, rotation could induce spin-spin interactions between the two black holes, which may affect their stability properties.

Even without rotation, we have partial results suggesting that, at least for a large portion of the moduli space, the solutions we have constructed are stable. When the black holes are far apart, the solutions are similar, at least locally, to the hovering black hole solutions presented in \cite{Horowitz:2014gva}, which were shown using a prove approximation to possess stable orbits for arbitrarily small extremal black holes. Furthermore, one can analyse the spectrum of the linearized Euclidean Einstein-DeTurck-Maxwell equations around the solutions. This linear operator must be computed as part of the numerical method regardless. For transverse-traceless perturbations of an Einstein-Maxwell solution, and for a Maxwell field with our degree of freedom, the spectrum coincides with that of the perturbed Einstein-Maxwell operator restricted to static, axisymmetric modes. Our analysis shows that all our solutions exhibit a single negative mode, corresponding to the standard thermodynamic instability of charged black holes in the grand-canonical ensemble. We find no evidence of additional unstable modes. Moreover, in four spacetime dimensions, this mode approaches those identified in \cite{Monteiro:2008wr}, which corresponds to a known stable solution.

The solutions described in this letter can be S-dualized into magnetically charged binaries. These binaries, in turn, could serve as the starting point for constructing a traversable wormhole, as demonstrated in Ref.~\cite{Maldacena:2018gjk}. The advantage of this approach is that the initial binary configuration exists as a static solution and may indeed prove to be stable. We leave this construction for future work.

Although the solutions presented in this letter possess no supersymmetry and are found at finite temperature, one might wonder whether supersymmetric solutions could exist. Given that the boundary chemical potential is necessarily nontrivial, one would need to search for possible supersymmetric solutions with spatially modulated deformations, perhaps using the approach pioneered in Ref.~\cite{Gauntlett:2018vhk}.

It would also be of great interest to understand if the binary solutions ever dominate over the soliton or polarised black hole solutions of Ref.~\cite{Costa:2015gol}. However, as discussed above, a thermodynamic investigation is difficult, since it is not clear there is a well-defined notion of free energy. If the horizons have unequal local chemical potentials, then the solutions are \textit{not} in thermoelectric equilibrium, and so it seems there is no good choice of ensemble in which to compare with the soliton and single horizon solutions. 

Furthermore, these solutions with $\nu_{L}\neq \nu_{R}$ become unstable under $\mathcal{O}(N^{-2})$ corrections since by the exchange of different electric charges via Hawking radiation Planckian-sized black holes, the system will eventually reaches global equilibrium. However, this process occurs over very long time scales.


This is perhaps akin to many-body localization, where systems undergo phases that exhibit local equilibrium but on much longer time scales eventually reach global equilibrium. Local equilibrium is usually associated with the existence of almost conserved quantities that make the system almost integrable, and that are responsible for delaying global thermalisation. In our case, one can conjecture that these local conserved quantities are the electric charges of each individual black hole \footnote{We would like to thank Sean~Hartnoll for the discussions pertaining to this paragraph.}. Though here there are only two black holes, this resemblance to many-body localization would become more manifest in systems with many disconnected black hole horizons; the existence of the binary solutions of this letter suggests strongly that such configurations will also exist. Indeed, even with the same $\ell=1$ boundary profile, one could envisage a static, axisymmetric configuration comprised of many black holes with alternating charges situated in a line along the $\partial_\phi$ axis of symmetry.

The solutions found in this Letter, let alone any of
these further conjectured solutions, constitute a plethora of multi-black-hole solutions in AdS. It would be extremely
desirable to better understand the states of the CFT which
are dual to these configurations, and in particular the CFT
interpretation of the extra parameters which arise. 

\paragraph*{Acknowledgements.}
We would like to thank \'Oscar~Dias, Gary~Horowitz, Maciej~Kolanowski and Grant~Remmen for reading an earlier version of this letter and for providing critical comments. W.~B. was supported by an STFC studentship, ST/S505298/1, and a Vice Chancellor's award from the University of Cambridge, and J. E. S. has been partially supported by STFC consolidated grants ST/T000694/1 and ST/X000664/1.
\newpage
\bibliography{papers}

\onecolumngrid
\clearpage
\appendix

\counterwithin*{equation}{section}
\renewcommand\theequation{\thesection\arabic{equation}}
\section{Obtaining the $\mathbb{Z}_2$ symmetric, $\ell = 1$ solution }\label{app:num}
We seek to find a solution to the Einstein-Maxwell equations which contains two black hole horizons and has the asymptotic structure of anti-de Sitter space. It is in theory possible to write down a global \textit{Ansatz} for such a solution, however, this is very difficult and not necessary. Instead the approach will be to define two different \textit{Ansätze}, one of which will be used to calculate the metric in a region near the black holes and the other in the asymptotic region.

These \textit{Ansätze} will be based on other solutions which share properties of the desired solution in the relevant region. Near the black holes, the cosmological constant has a smaller effect, so we will make use of an exact binary black hole solution with zero cosmological constant, the \textit{Israel-Khan} solution \cite{Israel:1964a}. On the other hand, the presence of the cosmological constant has much larger effect on the asymptotic structure than the presence of black holes in the bulk, and so in the asymptotic region we will instead base our \textit{Ansatz} on the metric of empty global AdS.

These two metrics will also be of great use when designing a reference metric which satisfies all the required boundary conditions. Let us now briefly review these two metrics before explaining how they are used to define the \textit{Ansätze} and the reference metric.
\subsection{The Israel-Khan Solution}
\begin{figure}
    \centering
    \includegraphics[width=0.6\textwidth]{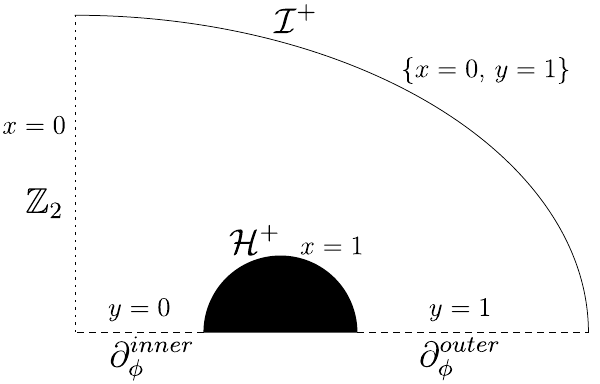}
    \caption{A sketch of the Israel-Khan spacetime. There is a plane of reflective symmetry at $x=0$, shown as a dotted line, and a black hole horizon at $x=1$. The $y=0$ and $y=1$ boundaries, shown as dashed lines, constitute the $\partial_\phi$ axis of rotation on the inside and outside of the binary system, respectively. The conical singularity is situated along the $y=0$ axis for the Israel-Khan metric. The ring-like $\{x,y\}$ coordinates are singular at infinity, with the boundary being described by the single coordinate point, $\{x = 0, y=1\}$.}
    \label{fig:IKsketch}
\end{figure}

The Israel-Khan solution is an exact solution to the vacuum Einstein equation in four dimensions with $\Lambda = 0$. It contains two black holes which are held apart by a conical strut. This solution can be written in ring-like coordinates, $\{x,y\}$, with the metric being given by
\begin{equation}
    \diff s^2_{IK} = - \Delta_x m_{xy}^2 (1-x^2)^2\diff t^2 + \frac{\lambda^2}{m_{xy}^2 \Delta_{xy}^2}\left(w_y^2\left(\frac{4 \diff x^2}{(2-x^2)\Delta_x}+\frac{4 \diff y^2}{(2-y^2)\Delta_y}\right) + y^2(2-y^2)(1-y^2)^2 \diff \phi^2\right),
\end{equation}
where
\begin{align}
    &\Delta_x(x) = 1-k^2 x^2(2-x^2), \quad\quad\quad \;\;\;
    \Delta_y(y) = 1-(1 - k^2)y^2(2-y^2),  \nonumber \\    & \Delta_{xy}(x,y) = (1-y^2)^2 + k^2x^2(2-x^2)y^2(2-y^2), \quad\quad\quad w_y(y) = \frac{k}{(1+k)^2}\left(1+\sqrt{\Delta_y(y)}\right)^2,\nonumber\\
    &m_{xy}(x,y) = \frac{k\left(1-(1-k)y^2(2-y^2)+\sqrt{\Delta_y(y)}\right)}{(1-k)\Delta_x(x)\left(1-y^2\right)^2+\left(k+\sqrt{\Delta_y(y)}\right)\left(\Delta_x(x)+(1-k)\left(\sqrt{\Delta_{xy}(x,y)}-1\right)\right)}.
\end{align}
The space of solutions is parameterised by $k \in (0,1)$ and $\lambda \in (0, \infty)$ with the temperature of each of the two horizons being given by
\begin{equation}
    T_H = \inv{2\pi}\frac{k(1+k)}{4\lambda (1-k)} .
\end{equation}
The Israel-Khan solution is static and axisymmetric with respect to the Killing vector fields $\partial_t$ and $\partial_\phi$, respectively. The other two coordinates lie in the range $x \in (-1,1)$ and $y \in (0,1)$. The horizons of the two black holes are situated at $x = \pm 1$, and there is a $\mathbb{Z}_2$ symmetry across the $x=0$ plane. Meanwhile, the $y=0$ and $y=1$ coordinate boundaries make up the $\partial_\phi$ axis of rotation, with the $y=0$ segment being \textit{inner axis}, the line between the two black hole horizons (at which the conical singularity is situated), and the $y=1$ segment constituting the \textit{outer axis}, the region of the $\partial_\phi$ axis between the horizons and infinity. The asymptotic boundary is situated the single coordinate point $(x,y)=(0,1)$ at which $\Delta_{xy}$ vanishes. Hence, there is a coordinate singularity at infinity in these coordinates. We have drawn a schematic diagram of a constant time slice of the Israel-Khan spacetime in Fig.~\ref{fig:IKsketch}.

If we were to use the Israel-Khan solution as a reference metric for the AdS binaries, they would inherit the conical singularity at $y=0$.
We, however, wish to obtain regular binary solutions, so let us generalise the Israel-Khan metric by multiplying the $\diff \phi^2$ term by a function $\Sigma(y)$ given by
\begin{equation}
    \Sigma(y) = 1-\alpha (1-y^2)^2.
\end{equation}
This gives us a family of metrics parameterised by $\alpha$:
\begin{equation}\label{eqn:warpedIK}
    \diff s^2_{IK;\Sigma} = - \Delta_x m_{xy}^2 (1-x^2)^2\diff t^2 + \frac{\lambda^2}{m_{xy}^2 \Delta_{xy}^2}\left\{w_y^2\left[\frac{4 \diff x^2}{(2-x^2)\Delta_x}+\frac{4 \diff y^2}{(2-y^2)\Delta_y}\right] + y^2(2-y^2)(1-y^2)^2 \Sigma(y) \diff \phi^2\right\},
\end{equation}
By adjusting the new parameter, $\alpha$, we can remove the conical singularity at $y=0$. In particular, if we take
\begin{equation}\label{eqn:alpha}
    \alpha = \frac{(1-k)^2(1+6k+k^2)}{(1+k)^4},
\end{equation}
then the metric, $\diff s^2_{IK;\Sigma}$, contains no conical singularities. Let us name the metric given by $\diff s^2_{IK;\Sigma}$ for this value of $\alpha$ the \textit{warped Israel-Khan metric}. Of course, $\diff s^2_{IK;\Sigma}$ is only a solution to the Einstein equation when $\alpha = 0$, but we will find that using the value of $\alpha$ which removes the conical singularity to be far more useful.
\subsection{Empty AdS}
Global anti-de Sitter in four dimensions is often written as
\begin{equation}\label{eqn:AdS}
    \diff s^2_{AdS} = -\left(1+\frac{R^2}{\ell_4^2}\right)\diff t^2+ \left(1+\frac{R^2}{\ell_4^2}\right)^{-1} \diff R^2+R^2 \left(\diff\theta^2 +\sin^2\theta \diff \phi^2\right),
\end{equation}
where $\ell_4$ is the AdS radius. For our uses, we will take new coordinates
\begin{equation}
    R = \frac{r}{1-r^2}, \quad \sin\theta = 1-\xi^2,
\end{equation}
so that the metric is given by
\begin{equation}\label{eqn:AdS_app}
    \diff s^2_{AdS} = \inv{(1-r^2)^2}\left\{-g(r) \diff t^2 + \frac{(1+r^2)^2}{g(r)} \diff r^2 + r^2 \left[\frac{4 \diff\xi^2}{2-\xi^2}+(1-\xi^2)^2 \diff \phi^2\right]\right\},
\end{equation}
with 
\begin{equation}
    g(r):= \frac{r^2}{\ell_4^2}+(1-r^2)^2.
\end{equation}
The radial coordinate transformation is necessary in order to compactify the coordinate, $r\in (0,1)$ with $r = 0$ being the origin, and $r=1$ the conformal boundary, which is topologically $\mathbb{R}_t\times S^2$. The transformation of the angular coordinate removes trigonometric functions from the metric which speeds up the numerics considerably. The $\partial_\phi$ axis is given by $\xi = \pm 1$, and there is a $\mathbb{Z}_2$ symmetry across $\xi = 0$.
\subsection{The \textit{Ansätze}}
As described above, we will have two \textit{Ansätze}, one for the region near the horizons based on the warped Israel-Khan metric and one for the asymptotic region based on empty global AdS. In each case, we will take the \textit{Ansatz} to be the most general deformation of the relevant metric which still satisfies all our symmetry assumptions. The binary solution in AdS will also be charged, and so we need also to define an \textit{Ansatz} for the Maxwell field. We will assume they are electrically charged, so that the vector potential satisfies $A_a \propto (\diff t)_a$.
\paragraph{The inner Ansatz.}
We define the \textit{Ansatz} in the region near the black holes in the $\{x,y\}$ coordinates of the Israel-Khan solution. We simply add unknown functions multiplying each metric component of the warped Israel-Khan metric, as well as adding an unknown $\diff x \diff y$ cross-term, yielding
\begin{subequations}
\begin{align}
    \diff s^2_{inner} = - \Delta_x m_{xy}^2 (1-x^2)^2 \mathcal{T}_{in}\diff t^2 + \frac{\lambda^2}{m_{xy}^2 \Delta_{xy}^2}\Bigg[w_y^2&\left(\frac{4\,\mathcal C_{in} \diff x^2}{(2-x^2)\Delta_x}+\frac{4\,\mathcal B_{in} }{(2-y^2)\Delta_y}\left(\diff y - \mathcal F_{in} \diff x\right)^2\right) \nonumber \\ &+ y^2(2-y^2)(1-y^2)^2 \Sigma(y) \,\mathcal S_{in} \diff \phi^2\Bigg].
\end{align}
For the gauge field, we simply take
\begin{equation}
    A = \mathcal A_{in}\diff t.
\end{equation}
\end{subequations}
Hence, we have six unknown functions in the in region that we have to solve for, $\{\mathcal{T}_{in},\, \mathcal{C}_{in},\, \mathcal{B}_{in},\, \mathcal{S}_{in},\, \mathcal{F}_{in},\, \mathcal{A}_{in}\}$, each of which depend on $\{x, y\}$. Note that by taking $\mathcal{T}_{in} = \mathcal{C}_{in} = \mathcal{B}_{in}=  \mathcal{S}_{in} =1$ and $\mathcal{F}_{in} = 0$, we obtain $\diff s^2_{IK;\Sigma}$, which we can then either take to be the Israel-Khan metric by setting $\alpha = 0$ or instead eradicate the conical singularity by taking $\alpha$ to be as given in (\ref{eqn:alpha}).

Though the whole spacetime is given by $x \in (-1,1)$, we will initially assume a $\mathbb{Z}_2$ symmetry at $x=0$ for the bulk solutions, and so we will focus on the $x \in (0,1)$ region, setting boundary conditions to enforce this symmetry.
\paragraph{The outer Ansatz.}
In the outer region we build the \textit{Ansatz} by considering a general deformation of the global AdS metric, whilst still respecting the desired symmetries. We take
\begin{subequations}
\begin{align}\label{eqn:out}
    \diff s^2_{outer} = \inv{(1-r^2)^2}\Bigg[&-g(r)\mathcal{T}_{out}\diff t^2 + \frac{(1+r^2)^2}{g(r)}\mathcal{C}_{out} \diff r^2  \nonumber \\ &+ r^2 \left(\frac{4 \mathcal{B}_{out}}{2-\xi^2}\left(\diff\xi - \mathcal{F}_{out} \diff r\right)^2+(1-\xi^2)^2 \mathcal{S}_{out}\diff \phi^2\right)\Bigg],
\end{align}
and for the gauge field
\begin{equation}
    A = \mathcal{A}_{out} \diff t.
\end{equation}
\end{subequations}
This time the unknown functions $\{\mathcal{T}_{out},\, \mathcal{C}_{out},\, \mathcal{B}_{out},\, \mathcal{S}_{out},\, \mathcal{F}_{out},\, \mathcal{A}_{out}\}$ depend upon the coordinates $\{r, \xi\}$. The global AdS metric is obtained by setting $\mathcal{T}_{out} = \mathcal{C}_{out} = \mathcal{B}_{out}=  \mathcal{S}_{out} =1$ and $\mathcal{F}_{out} = 0$.
\subsection{Coordinate Transformations}
We have a large amount of freedom in interpolating between the two coordinate domains. Firstly we can choose the position and shape of the interface between the two regions. In our case, we use a constant $r$ surface, $r = r_0$, as the interface. 
\begin{figure}[tb!]
    \centering
    \includegraphics[width=0.99\textwidth]{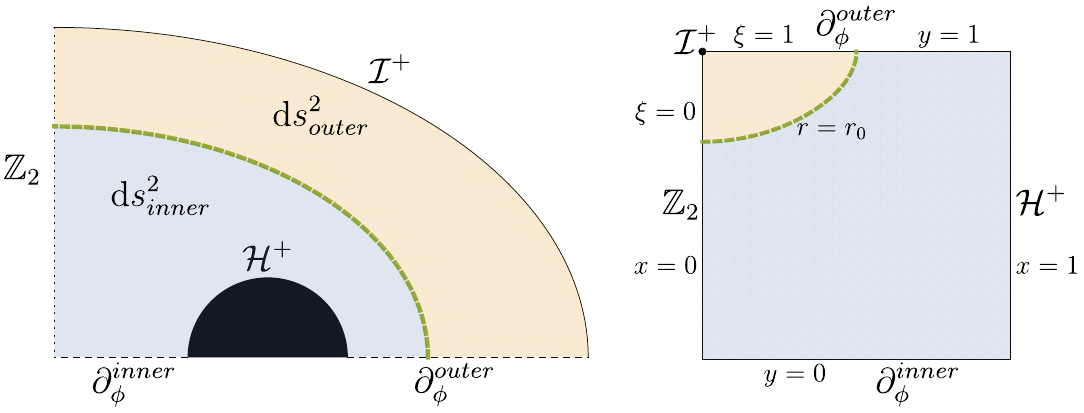}
    \caption{On the left we show the schematic drawing of the coordinate domain of the $\mathbb{Z}_2$-symmetric solutions. The sketch is split into the inner region, in blue, where the inner \textit{Ansatz} written in $\{x,y\}$ coordinates is taken and the outer region, in orange, where the outer \textit{Ansatz} in $\{r,\xi\}$ coordinates is used instead. On the right hand side we plot these region in the $(x,y)$ plane. Infinity is the point $(x,y)=(0,1)$ at which $r=1$. Near infinity, constant $r$ surfaces are curves from $x=0$, at which $\xi=0$, to $y=1$, at which $\xi =1$. The interface between the inner and outer regions is chosen to be such a constant $r$ slice, shown as a dashed green curve.}
    \label{fig:xyrxi_trans}
\end{figure}
Moreover, we have a further freedom in defining the coordinate transformation between the inner $\{x,y\}$ coordinates and the outer $\{r,\xi\}$ coordinates.

We firstly want to ensure that the outer $\partial_\phi$ axes and the $\mathbb{Z}_2$ reflection plane are located in the same positions at this interface, which requires that $\xi = 0 \iff x = 0$, and $\xi = 1 \iff y = 1$. Secondly, given a transformation, one can equate the \textit{Ansätze} at the interface, $r=r_0$, and thereby write each of $\{\mathcal{T}_{in},\, \mathcal{C}_{in},\, \mathcal{B}_{in},\, \mathcal{S}_{in},\, \mathcal{F}_{in},\, \mathcal{A}_{in}\}$ in terms of $\{\mathcal{T}_{out},\, \mathcal{C}_{out},\, \mathcal{B}_{out},\, \mathcal{S}_{out},\, \mathcal{F}_{out},\, \mathcal{A}_{out}\}$, and \textit{vice versa}. The key is to ensure that none of these relationships diverges at any point on $r=r_0$, including at the $\partial_\phi$ axis and the $\mathbb{Z}_2$ reflection plane. However, importantly, this need not be the case globally, only in the region in the vicinity of the interface of the two regions. Therefore, for example, the $r$ and $\xi$ coordinates need not be well defined at the horizons of the black holes or at the inner axis, both of which lie deep within the inner region.

The coordinate transformation we used is given by
\begin{equation}\label{eqn:xytorxi}
    x = (1-r)\xi\sqrt{2-\xi^2}, \quad y = \sqrt{1-(1-r)(1-\xi^2)},
\end{equation}
or, inversely,
\begin{equation}\label{eqn:rxitoxy}
    r = 1-\sqrt{x^2+(1-y^2)^2}, \quad \xi = \left(1-\frac{1-y^2}{\sqrt{1+x^2-y^2(2-y^2)}}\right)^{1/2}.
\end{equation}
In effect, $\{r, \xi\}$ can be thought of as a form of polar coordinates about the point $(x,y)=(0,1)$, which recall is the position of infinity for the Israel-Khan spacetime, taken such that $r = 1$ corresponds to this point.

Having taken this choice of coordinate transformation, it will be the boundary conditions  that we set on the boundary between the inner and outer regions at the $r=r_0$ interface which will ensure that the metric is smooth in the vicinity of the $r = r_0$. On the left hand side of Fig.~\ref{fig:xyrxi_trans} we give a schematic diagram of the domain of the $\mathbb{Z}_2$-symmetric binary solution, with the inner and outer regions in blue and orange, respectively. On the right of the figure, we plot these regions in the $(x,y)$ plane. The outer region is situated between infinity, at $(x,y)=(0,1)$ (or, equivalently, at $r=1$), and a constant $r$ slice, $r=r_0$.
\subsection{Boundary Conditions}
Let us consider the boundaries of our spacetime, discerning whether each of them lies within the outer or inner regions entirely, or straddles them both. We have the conformal boundary, which lies entirely in the outer region. We have the horizon at $x=1$ and the inner axis at $y=0$, which both lie solely in the inner region. And finally we have the $\mathbb{Z}_2$ reflection plane and the outer axis, which are present in both regions. Let us first deal with the boundary conditions for the inner and outer region before turning our attention to the boundary conditions we must set at the interface of the regions.
\paragraph{Inner boundary conditions.}
In the region near the black holes in which we use the ring-like $\{x,y\}$ coordinates we have a pentagonal domain. One boundary is the patching boundary which we will deal with momentarily. The remaining boundaries are dealt with as follows:
\begin{itemize}
    \item \underline{The $\mathbb{Z}_2$ reflection plane at $x = 0$}. Here we want the metric to be even and the gauge potential to be odd. Hence, we enforce at $x = 0$ that
    \begin{equation}
         0 = \pder[\mathcal{T}_{in}]{x} = \pder[\mathcal{C}_{in}]{x} =
        \pder[\mathcal{B}_{in}]{x} =
        \pder[\mathcal{S}_{in}]{x}  = 
         \mathcal{F}_{in}  = \mathcal{A}_{in}.
    \end{equation}
    \item \underline{The horizon at $x = 1$}. The metric functions must be regular across the horizon. Moreover, since the horizon is a Killing horizon, generated by $\partial_t$, the gauge potential must be constant on the horizon. Interestingly though, we find that this gauge potential need not be zero and indeed can take a range of values for a given boundary profile. Hence, we set at $x=1$
    \begin{equation}\label{eqn:BCx1}
        0 = \pder[\mathcal{T}_{in}]{x} = \pder[\mathcal{C}_{in}]{x} =
        \pder[\mathcal{B}_{in}]{x} =
        \pder[\mathcal{S}_{in}]{x}  = 
         \mathcal{F}_{in}, \quad \mathcal{A}_{in}  = \nu_R,
    \end{equation}
    where $\nu_R$ will be an extra parameter of the solution space, with the subscript $R$ denoting that it is the value of the vector potential on the right-hand horizon. This parameter is discussed in detail in the main text. Due to the fact the gauge potential is odd across $x=0$, the gauge potential at the left-hand horizon, which we denote as $\nu_L$ will have the opposite sign: $\nu_L = -\nu_R$.
    \item \underline{The inner $\partial_\phi$ axis at $y=0$} and \underline{the outer  $\partial_\phi$ axis at $y=1$}. At both of these boundaries, we take the boundary conditions
    \begin{equation}
        0 = \pder[\mathcal{T}_{in}]{y} = \pder[\mathcal{C}_{in}]{y} =
        \pder[\mathcal{B}_{in}]{y} =
        \pder[\mathcal{S}_{in}]{y}  = 
         \mathcal{F}_{in} = \pder[\mathcal{A}_{in}]{y}.
    \end{equation}
    Moreover, in order for there to be no conical singularity along this axis, we require that $\mathcal{B}_{in} = \mathcal{S}_{in}$ at these boundaries. This can be set as a boundary condition in place of one of the above, but even without doing this it will be enforced by the bulk equations of motion so long as the reference metric also has no conical singularities.
\end{itemize}
\paragraph{Outer boundary conditions.}
In the asymptotic region we use the $\{r,\xi\}$ coordinates, and the integration domain is rectangular. Once again, one boundary is the patching boundary to the inner region, and the other three are given by:
\begin{itemize}
    \item \underline{The $\mathbb{Z}_2$ reflection plane at $\xi = 0$}. Across this plane, the metric is even and the gauge field is odd. We set
    \begin{equation}
        0 = \pder[\mathcal{T}_{out}]{\xi} = \pder[\mathcal{C}_{out}]{\xi} =
        \pder[\mathcal{B}_{out}]{\xi} =
        \pder[\mathcal{S}_{out}]{\xi}  = 
         \mathcal{F}_{out} = \mathcal{A}_{out}.
    \end{equation}
    \item \underline{The outer $\partial_\phi$ axis at $\xi=1$}.  Here, we set
    \begin{equation}
        0 = \pder[\mathcal{T}_{out}]{\xi} = \pder[\mathcal{C}_{out}]{\xi} =
        \pder[\mathcal{B}_{out}]{\xi} =
        \pder[\mathcal{S}_{out}]{\xi}  = 
         \mathcal{F}_{out} =
         \pder[\mathcal{A}_{out}]{\xi}.
    \end{equation}
    Likewise to the discussion above regarding the $\partial_\phi$ axis of rotation in the inner coordinates, the absence of a conical singularity requires that $\mathcal{B}_{out} = \mathcal{S}_{out}$ at this boundary, though this again will be enforced by the equations of motion.
    \item \underline{The conformal boundary at $r = 1$}. Here we set Dirichlet boundary conditions. Holographically these boundary conditions correspond to the choice of metric on which the CFT will live and the choice of the background chemical potential with which we are deforming the CFT. In this case, we take
    \begin{equation}
        \mathcal{T}_{out} = \mathcal{C}_{out} = \mathcal{B}_{out} = \mathcal{S}_{out} = 1, \quad \mathcal{F}_{out} = 0, \quad \mathcal{A}_{out} = \mu_1\, \xi \sqrt{2-\xi^2}.
    \end{equation}
    These enforce that the spacetime is asymptotically AdS, having the same asymptotic structure as (\ref{eqn:AdS}). Meanwhile the boundary condition for the Maxwell field means that in the boundary theory we are exciting an $\ell = 1$ mode of the background electric field (note that $\xi \sqrt{2-\xi^2} = \cos \theta$ if we transform back to our familiar angular variable, $\theta$). The parameter, $\mu_1$, determines the magnitude of the enforced chemical potential.
\end{itemize}
\paragraph{Patching boundary conditions.}
Now we come to the tricky issue of the patching boundary conditions. In essence we require that the metric is \textit{continuous} and \textit{differentiable} at this interface, which is chosen to be a constant $r$ slice, $r=r_0$. 

In order to do this we can write the metric in either of the patches locally near the interface in the coordinates of the other patch by using the coordinate transformations given by (\ref{eqn:xytorxi}) and (\ref{eqn:rxitoxy}). Equating the two metrics at $r=r_0$ gives expressions for $\{\mathcal{T}_{in},\, \mathcal{C}_{in},\, \mathcal{B}_{in},\, \mathcal{S}_{in},\, \mathcal{F}_{in},\, \mathcal{A}_{in}\}$ in terms of $\{\mathcal{T}_{out},\, \mathcal{C}_{out},\, \mathcal{B}_{out},\, \mathcal{S}_{out},\, \mathcal{F}_{out},\, \mathcal{A}_{out}\}$, or \textit{vice versa}. 

On the boundary of one patch, we enforce these relationships as Dirichlet boundary conditions. This enforces continuity of the metric. As for the other patch, we take derivatives of the relationships between the inner and outer metric functions in the direction normal to the interface, and set these as boundary conditions. This then enforces that the metric is also differentiable at the interface of the patches.
\subsection{The Reference Metric}
Finally let us consider designing a reference metric, which must satisfy certain requirements in order to be suitable for use when solving the Einstein-DeTurck equation \eqref{eqn:EDT}. We need it to contain two black hole horizons, be regular and satisfy all the symmetry assumptions of the desired solution. Finally, we wish it to have the same asymptotic structure as the desired solution, that is, it must be asymptotically AdS. Each of these conditions is necessary (though not sufficient) to prevent the DeTurck method yielding a Ricci soliton, \textit{i.e.} a solution to the Einstein-DeTurck equation that is not a solution to the Einstein equation.

Importantly, however, we do \textit{not} require the reference metric to satisfy any field equations, though we do need it to be $C^2$, since the Einstein-DeTurck equations are second-order PDEs.

The warped Israel-Khan metric, \eqref{eqn:warpedIK}, satisfies all but the asymptotic condition. In particular, due to the warping factor included in the $\diff \phi^2 $ term, there is no conical singularity at $y=0$. On the other hand, the empty AdS metric, \eqref{eqn:AdS}, satisfies the asymptotic condition but not the condition regarding the presence of horizons. Thus, if we can design a metric that interpolates from the former to the latter as we move from the inner region to the outer region, we shall have a reference metric satisfying all our desired properties. 

This can be done in a number of ways. In \cite{Headrick:2009pv} a suitable reference metric is obtained by interpolating with an interpolation function with compact support only on a subregion of the spacetime, whilst in \cite{Figueras:2014dta} an interpolation function with support across the whole spacetime is used. In our case, we use a method most similar to the former. 

In the entire inner region, we take the reference metric to be equal to the warped Israel-Khan metric, \eqref{eqn:warpedIK}. Then we shall split the outer region, which recall is given by $r \in (r_0,1)$, into two sections with an interface, $r = r_1$, that we are free to choose with $r_0<r_1<1$. In the asymptotic region $r \in (r_1,1)$, we will take the reference metric to be the global AdS metric, \eqref{eqn:AdS_app}. In the intermediate region, $r \in (r_0,r_1)$, we take an interpolation between the warped Israel-Khan and empty AdS metrics, that is we have
\begin{equation}\label{eqn:int_ref}
    \Bar{\diff s}^2_{inter} = I(r)\diff s^2_{IK;\Sigma} + \left(1-I(r)\right) \diff s^2_{AdS},
\end{equation}
where $I(r)$ is an interpolating function which is equal to one at $r=r_0$ and vanishes at $r=r_1$ to sufficiently high order such that the reference metric is $C^2$. In order to obtain $\diff s^2_{IK;\Sigma}$ in the outer $\{r,\xi\}$ coordinates, we simply apply our coordinate transformations, \eqref{eqn:xytorxi}, which are regular for $r<r_1$. This was one benefit of using an intermediate region rather than interpolating the reference metrics across the whole of the outer region, \textit{i.e.} all the way from $r=r_0$ to $r=1$; we never needed to deal with difficulties arising due to the coordinate transformation becoming singular at infinity (though these difficulties are certainly not insurmountable). Moreover, we found that taking the reference metric in the asymptotic region to be exactly empty AdS eased the extraction of holographic quantities.

There is a great amount of choice in taking the interpolating function, $I(r)$. One could ensure that at $r=r_0$ and $r=r_1$ the function approaches one and zero, respectively, to all orders by using a non-analytic function based on, for example, a hyperbolic tangent function. However, this introduction of non-analyticities slows down the numerics and also can cause large gradients in the solutions. In fact, in order to obtain a $C^2$ reference metric, we only need the interpolating function to approach the desired values at the endpoints, $r_0$ and $r_1$, to second order. We choose the following interpolating function:
\begin{equation}
    I(r) = \left(\frac{r-r_1}{r_0-r_1}\right)^4\left(10 - 20 \left(\frac{r-r_1}{r_0-r_1}\right)^2 + 15\left(\frac{r-r_1}{r_0-r_1}\right)^4 - 4\left(\frac{r-r_1}{r_0-r_1}\right)^6\right),
\end{equation}
which, respectively, approaches one and zero at $r_0$ and $r_1$ up to third order in $r$.

This choice of reference metric is compatible with all the boundary conditions.
\section{Patching and Numerical Method}
We have two coordinate domains, the inner and outer regions, with coordinate systems $\{x,y\}$ and $\{r,\xi\}$, respectively. The inner region is also naturally pentagonal, and so to carry out numerics we split it into two separate rectangular patches. Moreover, we split the outer region into two patches, so that we can easily fix the reference metric to be the interpolating reference metric, \eqref{eqn:int_ref}, in an intermediate patch and the empty AdS metric in the asymptotic patch.

\begin{figure}[t!]
    \centering
    \includegraphics[width=0.6\textwidth]{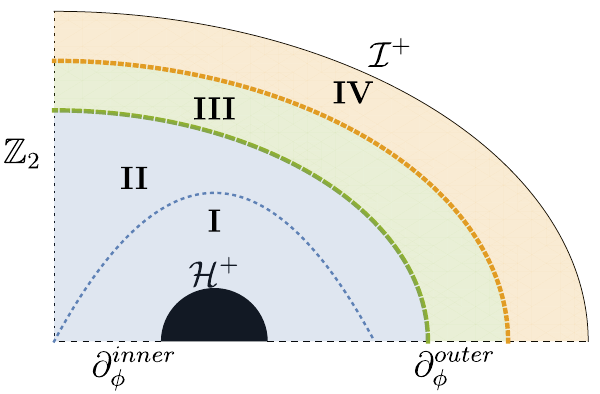}
    \caption{A schematic drawing of the coordinate domain, and the four patches used. In the inner region, shaded blue and composed of patches I and II, the ring-like $\{x,y\}$ coordinates are used, whereas in the outer region, which is shaded orange and made up of patches III and IV, the $\{r, \xi\}$ coordinates are used. The reference metric is given by the warped Israel-Khan metric in regions I and II and the empty AdS metric in region IV, whilst in region III it interpolates between these two metrics in a continuous and differentiable manner.}
    \label{fig:patches}
\end{figure}

In Fig. \ref{fig:patches}, we have sketched how the four patches are situated in the spacetime. Patch I and II (blue) constitute the inner region, where the ring-like $\{x,y\}$ coordinates are used and the reference metric is the warped Israel-Khan metric. 
Meanwhile, the reference metric is global AdS in region IV (orange) and so $\{r,\xi\}$ coordinates are used. The reference metric interpolates in a continuous and differentiable fashion from the warped Israel-Khan metric to global AdS in patch III (green). In order to define this interpolating reference metric it is necessary that both the $\{x,y\}$ and $\{r,\xi\}$ coordinate systems and the transformation between them are well-defined in this intermediate patch. Consequently, either the inner or outer \textit{Ansatz} could be used in this patch, but since its boundaries are constant $r$ and constant $\xi$ surfaces, it is most natural to use the outer \textit{Ansatz}.

There is a great deal of freedom when fixing the precise position of the boundaries between patches. We took the boundaries between patch I and II to be $x = x_0 y \sqrt{2-y^2}$, between patch II and III to be $r = r_0$, and between patch III and IV to be $r=r_1$, for fixed constants, $x_0,\,r_0$ and $r_1$. For the most part, we took these patching parameters to have the values $x_0=0.5,\, r_0=0.5,\, r_1=0.95$. One can also add extra patches in regions in which the solutions have large derivatives in order to speed up the numerical method and improve the accuracy of the solutions.

We discretize each of the patches with an $N \times N$ Chebyshev-Gauss-Lobatto grid and use transfinite interpolation and pseudospectral methods to approximate the PDEs by a large set of non-linear algebraic equations, which can then be solved iteratively with the Newton-Raphson method.

The main difficulty with this approach is to find a good seed which will converge to a solution. To alleviate this we the so-called $\delta$-trick, which is described in Sec.~VII.A of \cite{Dias:2015nua}. This allows one to artificially begin with equations that necessarily have a certain solution and from there slowly adapt the equations of motion towards those which we really are aiming to solve, \textit{i.e.} the Einstein-DeTurck and Maxwell equations, at each step solving intermediate equations. We also found it useful to vary $\alpha$, the parameter of the family of the warped Israel Khan metric, given in \eqref{eqn:warpedIK}, during this process, ensuring that at the end we land on the value of $\alpha$ given in \eqref{eqn:alpha} which ensures there are no conical singularities. In our case, we begin with equations that are automatically solved by the reference metric with $\alpha = 0$ and a vanishing Maxwell field.





\section{Generalising to the non-symmetric case and five dimensions}
\subsection{Removing the $\mathbb{Z}_2$ symmetry}
In the above, we assumed that the bulk solution was $\mathbb{Z}_2$ symmetric. Now let us drop this assumption to find more general solutions.

\begin{figure}[tb!]
    \centering
    \includegraphics[width=0.7\linewidth]{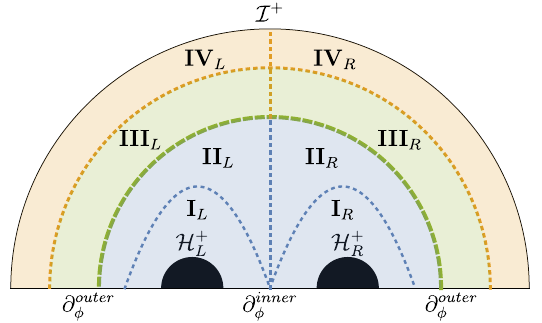}
    \caption{The coordinate domain for the non-symmetric solutions, split now into eight different patches. Similarly to the symmetric case, the reference metric is warped Israel-Khan near the black holes (blue), global AdS in the asymptotic region (orange) and interpolates between the two in the intermediate region (green).}
    \label{fig:patches_nosym}
\end{figure}

Without this symmetry, we cannot no longer use the $\xi = x=0$ plane as a boundary of the coordinate domain, and instead have to solve for the full range of the angular coordinates, \textit{i.e.} for $x \in (-1,+1)$ in the inner coordinate system and for $\xi \in (-1,+1)$ in the outer coordinate system. We can find a system of patches for this domain simply by doubling the patches from the symmetric case. Hence we have eight patches, as shown in Fig.~\ref{fig:patches_nosym}, which we have labelled with the Roman numerals, \textbf{I}-\textbf{IV}, and the subscripts $L$ and $R$ (denoting left and right, respectively) in order to keep touch with the patches in the symmetric case.

Note, that we need to set boundary conditions at both the right horizon, $\mathcal{H}^+_R$, at $x=+1$ and the left horizon, $\mathcal{H}^+_L$, at $x=-1$. These will both be of the form described in \eqref{eqn:BCx1}.

With the boundary potential taken (in the usual polar coordinate, $\theta$) to be $\mu(\theta) = \mu_1 \cos \theta$, we assumed in the method described in Appendix~\ref{app:num} that the vector potential inherited the odd symmetry of this boundary potential about $\theta = \pi/2$ in the bulk. Hence, we were implicitly setting $\nu_L = -\nu_R$. In general, though, we can set the value of the gauge potential to be two unrelated values at the two horizons:
\begin{equation}
    \mathcal{A}_{in}(-1,y) = \nu_L \quad\quad \text{and} \quad\quad \mathcal{A}_{in}(+1,y) = \nu_R.
\end{equation}
and break the boundary symmetry in the bulk by enforcing that $\nu_L \neq - \nu_R$. We were able to find such solutions by perturbing away from the symmetric solutions. 

We can also explictly break the $\mathbb{Z}_2$ symmetry of the boundary profile. For example, we were able to obtain solutions arising from a boundary profile in which an $\ell = 2$ mode was added to the $\ell = 1$ mode, given by
\begin{equation}
    \mu(\theta) = \mu_1 \cos \theta + \mu_2 \cos 2\theta.
\end{equation}

\subsection{Changing the number of dimensions}
The method to obtain solutions in five dimensions is almost identical to that described above for axisymmetric four-dimensional solutions. In each metric described, one need only replace the $\diff \phi^2$ by the metric of a two-sphere, $\diff \Omega^2 = \diff \psi^2 +\sin^2 \psi \diff \phi^2$. Hence rather than axisymmetry, the five-dimensional binaries possess an $SO(3)$ symmetry, with a two-sphere being preserved.

The key difficulty is finding a first solution in five dimensions. To do so, we used the $\delta$-trick once again, this time beginning with the equations of motion of the four dimensional problem. Specifically, if we let $E_{4d}$ and $E_{5d}$ denote the equations of motion of the four- and five-dimensional problems, respectively, we considered
\begin{equation}
    E(\delta) = \delta\, E_{4d} + (1-\delta)\,E_{5d}.
\end{equation}
Thus, a four-dimensional binary solution solves $E(1)=0$. We then slowly decreased $\delta$ to zero, solving $E(\delta)=0$ at each intermediate stage, using the previous solution as a seed, to obtain a five-dimensional binary solution, satisfying $E(0) = 0$. We found it necessary to move around the parameter space (specifically increasing $\nu_R$ or decreasing $\mu_1$) during the process as $\delta$ decreased in order for the equation, $E(\delta)=0$, to continue having a solution at each intermediate value of $\delta$. Once an initial five-dimensional binary solution is found, it is fairly easy to perturb from this starting point to find other solutions nearby in the parameter space.
\section{Extracting the holographic quantities}\label{app:holo}
Let us review how the holographic stress tensor and conserved current of the boundary CFT can be extracted from the bulk solutions, using the procedure of holographic renormalization \cite{deHaro:2000vlm}. We will focus on the case of the four-dimensional solutions, and we will set $\ell_4 = 1$, as we did in the main text. The holographic quantities of the five-dimensional bulk solutions can also be extracted, though in this case the procedure is much more complicated due to the presence of the conformal anomaly in odd bulk dimensions (see, for example, Appendix A of Ref. \cite{Dias:2022eyq} for explicit formulae to extract the holographic quantities in this case).

We consider the metric locally near the conformal boundary. This is defined in terms of the $\{r,\xi\}$ coordinates of the outer \textit{Ansatz}, given by \eqref{eqn:out}, and the boundary is the $r=1$ surface. We first expand the equations of motion and the requirement that the DeTurck vector vanishes order-by-order in $r$ around $r=1$, whilst also assuming the asymptotic boundary conditions. This local analysis restricts the form of the unknown metric functions near the boundary to a large degree:
\begin{subequations}\label{eqn:r=1expansion}
\begin{align}
    \mathcal{T}_{out}(r,\xi) &= 1+ \alpha_1(\xi)\left(1-r\right)^3 + \left(\frac{3}{2}\alpha_1(\xi) + 4\alpha_6(\xi)^2\right)\left(1-r\right)^4 +\gamma_1(\xi)\left(1-r\right)^{\inv{2}(3+\sqrt{33})} + \hdots \\
     \mathcal{C}_{out}(r,\xi) &= 1 +4\left((2-\xi^2)\mu'(\xi)-\alpha_6(\xi)^2\right)\left(1-r\right)^4 +\gamma_2(\xi)\left(1-r\right)^{\inv{2}(3+\sqrt{33})} + \hdots \\
     \mathcal{B}_{out}(r,\xi) &= 1+ \alpha_3(\xi)\left(1-r\right)^3 + \frac{3}{2}\alpha_3(\xi)\left(1-r\right)^4 +\gamma_1(\xi)\left(1-r\right)^{\inv{2}(3+\sqrt{33})} + \hdots \\
     \mathcal{S}_{out}(r,\xi) &= 1 -\left(\alpha_1(\xi)+\alpha_3(\xi)\right)\left(1-r\right)^3 \nonumber \\
     &\quad\quad-\left(4(2-\xi^2)\mu'(\xi)^2+\frac{3}{2}\left(\alpha_1(\xi)+\alpha_3(\xi)\right)\right)\left(1-r\right)^4 +\gamma_1(\xi)\left(1-r\right)^{\inv{2}(3+\sqrt{33})} + \hdots \\
     \mathcal{F}_{out}(r,\xi) &= \beta_5(\xi)\left(1-r\right)^4 +\frac{16}{3}(2-\xi^2)\mu'(\xi)\alpha_6(\xi)\left(1-r\right)^4 \log{(1-r)} + \hdots \\
     \mathcal{A}_{out}(r,\xi) &= \mu(\xi) + \alpha_6(\xi)(1-r) + \hdots
\end{align}
\end{subequations}
where all terms in the expansion (including the higher order terms denoted by $\hdots$) are determined in terms of the six unknown functions, $\{\alpha_1, \alpha_3, \alpha_6, \beta_5, \gamma_1, \gamma_2\}$. In order to determine these six functions though, one must solve the equations in the full spacetime, whilst requiring regularity deep within the bulk. Fortunately, we will see only the $\{\alpha_1,\alpha_3,\alpha_6\}$ functions arise in the holographic stress tensor and conserved current, and so, in particular, the coefficients of the non-analytic terms, $\gamma_1$ and $\gamma_2$, do not need to be calculated in the analysis. Moreover the asymptotic analysis also enforces that the $\alpha$ functions are related by \begin{equation}\label{eqn:Wardalpha}
    \alpha_3'(\xi) = \frac{2 \left(3
   \xi  \left(\alpha_1(\xi )+2  \alpha_3(\xi )\right) -8 \left(1-\xi ^2\right) \alpha_6(\xi ) \mu'(\xi )\right)}{3 \left(1-\xi
   ^2\right)}.
\end{equation}
Next we can transform to Fefferman-Graham gauge \cite{Fefferman:1985} near the conformal boundary, \textit{i.e.} we seek a coordinate transformation which locally near the boundary brings the metric and gauge potential into the form
\begin{subequations}
\begin{align}\label{eqn:FG}
    \diff s^2&=\inv{z^2}\left[\diff z^2+ \left(g^{(0)}_{\mu\nu}+g^{(2)}_{\mu\nu} z^2+ g^{(3)}_{\mu\nu}z^3\right)\diff x^\mu \diff x^\nu + \bigO{z^4}\right] \\
    A &= \left(A_{\mu}^{(0)} + A_\mu^{(1)}z + \bigO{z^2} \right) \diff x^\mu
\end{align}
\end{subequations}
where $g^{(0)}$ is metric of the Einstein static Universe and $A^{(0)}$ is the boundary chemical potential, and the Greek indices run over the coordinates which span the boundary. This can be achieved simply by taking
\begin{align}
    r &= 1 - \inv{2}z + \inv{8}z^2 - \inv{8}z^3 + \frac{7}{128}z^4.
\end{align}
Once the metric is in this gauge, the vacuum expectation values of the holographic stress tensor and the conserved current can be, respectively, read off as
\begin{equation}
    \langle T_{\mu\nu} \rangle  = \frac{3}{16\pi G_N} g^{(3)}_{\mu\nu},  \quad\quad \langle J^\mu \rangle = \inv{4\pi G_N} A^{(1)}_\mu.
\end{equation}
Using the expansion of the metric functions near the conformal boundary, given in \eqref{eqn:r=1expansion}, one finds that these holographic quantities are given in $\{t,\xi,\phi\}$ coordinates of the boundary metric by
\begin{subequations}
\begin{align}
    \langle T^\mu{}_\nu \rangle &= \frac{3}{128 \pi G_N}\text{diag}\left(\alpha_1, \alpha_3, -\alpha_1-\alpha_3\right) \\
    \langle J^\mu \rangle &=\inv{{8\pi G_N}} \left(\alpha_6,0,0\right).
\end{align}
\end{subequations}
Note that the stress tensor is traceless and the current is conserved, $D_\mu \langle J^\mu \rangle = 0$, where $D_\mu$ is the covariant derivative arising from the boundary metric. Moreover, due to the relation given in \eqref{eqn:Wardalpha}, the stress tensor and conserved current satisfy a Ward identity, given by
\begin{equation}
    D_\mu \langle T^\mu{}_\nu \rangle = F^{(0)}_{\mu \nu}\langle J^\mu \rangle,
\end{equation}
where $F^{(0)}$ is the field strength tensor associated to the boundary potential, $A^{(0)}$.

The total energy can be computed by an integral of the holographic stress tensor:
\begin{equation}
    E \equiv -\int_{\Sigma}\diff \bm{x} \sqrt{h}n^\mu k^\nu \langle T_{\mu\nu}\rangle.
\end{equation}
where $\Sigma$ is a Cauchy slice of the boundary geometry, which in our case is a $(d-2)$-dimensional sphere, with $h$ denoting the determinant of its induced metric. Meanwhile, $k^\mu$ is the stationary Killing vector field of the boundary geometry and $n^\mu$ the unit normal to $\Sigma$. 

\section{Convergence tests}\label{app:conv}
We need to check explicitly that solutions obtained are not Ricci solitons, which are solutions to the Einstein-DeTurck equation, \eqref{eqn:EDT}, with non-zero DeTurck vector, $\xi^a$, and hence are not solutions to the Einstein equation. Since we are solving the equations numerically with each patch being approximated by a discrete grid of resolution $N\times N$, the requirement is that the DeTurck vector vanishes in the continuum limit, $N \to \infty$.

\begin{figure}[b!]
    \centering
    \subfloat[$\mathbb{Z}_2$ solution with $T=1$, $\mu_1 = 1.5$ and $\nu_R = -0.45$\label{fig:conv4d} in 4d]{\includegraphics[width=0.43\textwidth]{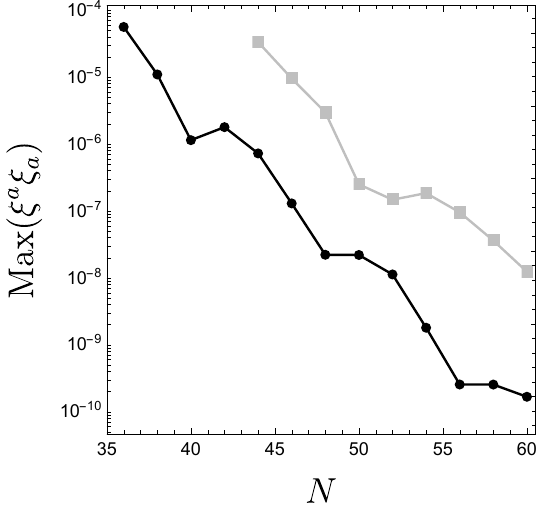}}\hfill
    \centering
    \subfloat[$\mathbb{Z}_2$ solution with $T=1$, $\mu_1 = 1.4$ and $\nu_R = 0$ in 5d\label{fig:conv5d}]{\includegraphics[width=0.43\textwidth]{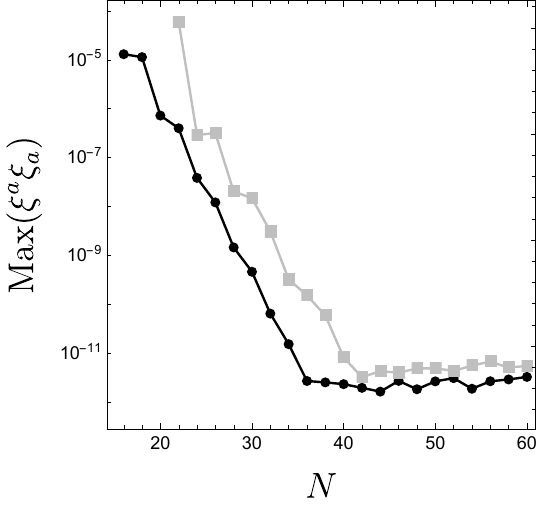}}
    \caption{The maximum value of the DeTurck norm for binary solutions obtained with a resolution in each patch of $N \times N$. The left corresponds to a four-dimensional solution and the right to a five-dimensional solution. In each case, two different solutions, differentiated by shade, are found for the values of the input parameters, either side of a turning point.}
    \label{fig:convergence}
\end{figure}

Fig.~\ref{fig:convergence} shows a log-plot the maximum value of the norm of the DeTurck vector across each lattice point of each patch for the numerical solutions obtained with different values of the resolution, $N$.
The left panel corresponds to four-dimensional, $\mathbb{Z}_2$-symmetric binary solutions with $T=1$, $\mu_1 = 1.5$ and $\nu_R = -0.45$, whilst the solutions in the right panel are five-dimensional, $\mathbb{Z}_2$-symmetric binary solutions with $T=1$, $\mu_1 = 1.4$ and $\nu_R = 0$.  In each case we were able to find two different solutions for the given input parameters, either side of a turning point. We have used different shades to differentiate between the solutions from the two different branches. The shades used in the left panel correspond with those used for the two different branches in Fig.~\ref{fig:turning} and Fig.~\ref{fig:SEQQ}.

The specific four-dimensional solution chosen has relatively large derivatives, explaining why a large number of points is needed to reduce the DeTurck vector. On the other hand, the chosen five-dimensional solution converges more quickly before hitting numerical precision at around $10^{-12}$. Both plots are consistent with power-law convergence, as expected for the DeTurck method.

\end{document}